\RequirePackage{lineno}
\documentclass[prd,onecolumn,groupedaddress,superscriptaddress,amsmath,amssymb]{revtex4}

\RequirePackage[colorlinks,citecolor=blue,urlcolor=blue,linkcolor=blue]{hyperref}

\usepackage{graphicx}
\usepackage{dcolumn}
\usepackage{bm}
\usepackage{amssymb}
\usepackage{enumerate}
\usepackage{lineno}
\usepackage{footnote}
\usepackage{doi} 

\newcommand\ee{e^+e^-}

\newcommand\pair{e^+e^-}

\def\address{\@ifstar{\address@star}%
  {\@ifnextchar[{\address@optarg}{\address@noptarg}}}

\begin{document}

\title{
\begin{center}
\end{center}
\vskip0.5cm
Search for pseudoscalar bosons decaying into $\pair$ pairs in the NA64 experiment at the CERN SPS.
}

\affiliation{\it Universit\"at Bonn, Helmholtz-Institut f\"ur Strahlen-und Kernphysik, 53115 Bonn, Germany} 
\affiliation{\it Joint Institute for Nuclear Research, 141980 Dubna, Russia}
\affiliation{\it Technische Universit\"at M\"unchen, Physik  Department, 85748 Garching, Germany}
\affiliation{\it CERN, European Organization for Nuclear Research, CH-1211 Geneva, Switzerland}
\affiliation{\it University of Illinois at Urbana Champaign, Urbana, 61801-3080 Illinois, USA}
\affiliation{\it UCL Departement of Physics and Astronomy, University College London, Gower St. London WC1E 6BT, United Kingdom}
\affiliation{\it Institute for Nuclear Research, 117312 Moscow, Russia}
\affiliation{\it P.N.Lebedev Physical Institute of the Russian Academy of Sciences, 119 991 Moscow, Russia}
\affiliation{\it Skobeltsyn Institute of Nuclear Physics, Lomonosov Moscow State University, 119991  Moscow, Russia}
\affiliation{\it Physics Department, University of Patras, 265 04 Patras, Greece} 
\affiliation{\it State Scientific Center of the Russian Federation Institute for High Energy Physics of National Research Center 'Kurchatov Institute' (IHEP), 142281 Protvino, Russia}
\affiliation{\it Departamento de Ciencias F\'{i}sicas, Universidad Andres Bello, Sazi\'{e} 2212, Piso 7, Santiago, Chile}
\affiliation{\it Tomsk Polytechnic University, 634050 Tomsk, Russia}
\affiliation{\it Tomsk State Pedagogical University, 634061 Tomsk, Russia}
\affiliation{\it Universidad T\'{e}cnica Federico Santa Mar\'{i}a, 2390123 Valpara\'{i}so, Chile}
\affiliation{\it ETH Z\"urich, Institute for Particle Physics and Astrophysics, CH-8093 Z\"urich, Switzerland}
\affiliation{\it Instituto de Fisica Corpuscular (CSIC/UV), Carrer del Catedrátic José Beltrán Martinez, 2, 46980 Paterna, Valencia}
\author{Yu.~M.~Andreev}\affiliation{\it Institute for Nuclear Research, 117312 Moscow, Russia}
\author{D.~Banerjee}\affiliation{\it CERN, European Organization for Nuclear Research, CH-1211 Geneva, Switzerland}
\author{J.~Bernhard}\affiliation{\it CERN, European Organization for Nuclear Research, CH-1211 Geneva, Switzerland}
\author{V.~E.~Burtsev}\affiliation{\it Joint Institute for Nuclear Research, 141980 Dubna, Russia}
\author{N.~Charitonidis}\affiliation{\it CERN, European Organization for Nuclear Research, CH-1211 Geneva, Switzerland}
\author{A.~G.~Chumakov}\affiliation{\it Tomsk Polytechnic University, 634050 Tomsk, Russia}\affiliation{\it Tomsk State Pedagogical University, 634061 Tomsk, Russia}
\author{D.~Cooke}\affiliation{\it UCL Departement of Physics and Astronomy, University College London, Gower St. London WC1E 6BT, United Kingdom}
\author{P.~Crivelli}\affiliation{\it ETH Z\"urich, Institute for Particle Physics and Astrophysics, CH-8093 Z\"urich, Switzerland}
\author{E.~Depero}\affiliation{\it ETH Z\"urich, Institute for Particle Physics and Astrophysics, CH-8093 Z\"urich, Switzerland}
\author{A.~V.~Dermenev}\affiliation{\it Institute for Nuclear Research, 117312 Moscow, Russia}
\author{S.~V.~Donskov}\affiliation{\it State Scientific Center of the Russian Federation Institute for High Energy Physics of National Research Center 'Kurchatov Institute' (IHEP), 142281 Protvino, Russia}
\author{R.~R.~Dusaev}\affiliation{\it Tomsk Polytechnic University, 634050 Tomsk, Russia}
\author{T.~Enik}\affiliation{\it  Joint Institute for Nuclear Research, 141980 Dubna, Russia}
\author{A.~Feshchenko}\affiliation{\it  Joint Institute for Nuclear Research, 141980 Dubna, Russia}
\author{V.~N.~Frolov}\affiliation{\it  Joint Institute for Nuclear Research, 141980 Dubna, Russia}
\author{A.~Gardikiotis}\affiliation{\it Physics Department, University of Patras, 265 04 Patras, Greece}
\author{S.~G.~Gerassimov }\affiliation{\it Technische Universit\"at M\"unchen, Physik  Department, 85748 Garching, Germany}\affiliation{\it P.N.Lebedev Physical Institute of the Russian Academy of Sciences, 119 991 Moscow, Russia}
\author{S.~N.~Gninenko}\affiliation{\it Institute for Nuclear Research, 117312 Moscow, Russia}
\author{M.~H\"osgen}\affiliation{\it Universit\"at Bonn, Helmholtz-Institut f\"ur Strahlen-und Kernphysik, 53115 Bonn, Germany}
\author{M.~Jeckel}\affiliation{\it CERN, European Organization for Nuclear Research, CH-1211 Geneva, Switzerland}
\author{V.~A.~Kachanov}\affiliation{\it State Scientific Center of the Russian Federation Institute for High Energy Physics of National Research Center 'Kurchatov Institute' (IHEP), 142281 Protvino, Russia}
\author{A.~E.~Karneyeu}\affiliation{\it Institute for Nuclear Research, 117312 Moscow, Russia}
\author{G.~Kekelidze}\affiliation{\it  Joint Institute for Nuclear Research, 141980 Dubna, Russia}
\author{B.~Ketzer}\affiliation{\it Universit\"at Bonn, Helmholtz-Institut f\"ur Strahlen-und Kernphysik, 53115 Bonn, Germany}
\author{D.~V.~Kirpichnikov}\affiliation{\it Institute for Nuclear Research, 117312 Moscow, Russia}
\author{M.~M.~Kirsanov}\affiliation{\it Institute for Nuclear Research, 117312 Moscow, Russia}
\author{V.~N.~Kolosov}\affiliation{\it State Scientific Center of the Russian Federation Institute for High Energy Physics of National Research Center 'Kurchatov Institute' (IHEP), 142281 Protvino, Russia}
\author{I.~V.~Konorov}\affiliation{\it Technische Universit\"at M\"unchen, Physik  Department, 85748 Garching, Germany}\affiliation{\it P.N.Lebedev Physical Institute of the Russian Academy of Sciences, 119 991 Moscow, Russia}
\author{S.~G.~Kovalenko}\affiliation{\it Departamento de Ciencias F\'{i}sicas, Universidad Andres Bello, Sazi\'{e} 2212, Piso 7, Santiago, Chile}
\author{V.~A.~Kramarenko}\affiliation{\it  Joint Institute for Nuclear Research, 141980 Dubna, Russia}\affiliation{\it Skobeltsyn Institute of Nuclear Physics, Lomonosov Moscow State University, 119991  Moscow, Russia}
\author{L.~V.~Kravchuk}\affiliation{\it Institute for Nuclear Research, 117312 Moscow, Russia}
\author{ N.~V.~Krasnikov}\affiliation{\it  Joint Institute for Nuclear Research, 141980 Dubna, Russia}\affiliation{\it Institute for Nuclear Research, 117312 Moscow, Russia}
\author{S.~V.~Kuleshov}\affiliation{\it Departamento de Ciencias F\'{i}sicas, Universidad Andres Bello, Sazi\'{e} 2212, Piso 7, Santiago, Chile}
\author{V.~E.~Lyubovitskij}\affiliation{\it Tomsk Polytechnic University, 634050 Tomsk, Russia}\affiliation{\it Tomsk State Pedagogical University, 634061 Tomsk, Russia}\affiliation{\it Universidad T\'{e}cnica Federico Santa Mar\'{i}a, 2390123 Valpara\'{i}so, Chile}
\author{V.~Lysan}\affiliation{\it  Joint Institute for Nuclear Research, 141980 Dubna, Russia}
\author{V.~A.~Matveev}\affiliation{\it  Joint Institute for Nuclear Research, 141980 Dubna, Russia}
\author{Yu.~V.~Mikhailov}\affiliation{\it State Scientific Center of the Russian Federation Institute for High Energy Physics of National Research Center 'Kurchatov Institute' (IHEP), 142281 Protvino, Russia}
\author{L.~Molina Bueno}\affiliation{\it ETH Z\"urich, Institute for Particle Physics and Astrophysics, CH-8093 Z\"urich, Switzerland}\affiliation{\it Instituto de Fisica Corpuscular (CSIC/UV), Carrer del Catedrátic José Beltrán Martinez, 2, 46980 Paterna, Valencia}
\author{D.~V.~Peshekhonov}\affiliation{\it  Joint Institute for Nuclear Research, 141980 Dubna, Russia}
\author{V.~A.~Polyakov}\affiliation{\it State Scientific Center of the Russian Federation Institute for High Energy Physics of National Research Center 'Kurchatov Institute' (IHEP), 142281 Protvino, Russia}
\author{B.~Radics}\affiliation{\it ETH Z\"urich, Institute for Particle Physics and Astrophysics, CH-8093 Z\"urich, Switzerland}
\author{R.~Rojas}\affiliation{\it Universidad T\'{e}cnica Federico Santa Mar\'{i}a, 2390123 Valpara\'{i}so, Chile}
\author{A.~Rubbia}\affiliation{\it ETH Z\"urich, Institute for Particle Physics and Astrophysics, CH-8093 Z\"urich, Switzerland}
\author{V.~D.~Samoylenko}\affiliation{\it State Scientific Center of the Russian Federation Institute for High Energy Physics of National Research Center 'Kurchatov Institute' (IHEP), 142281 Protvino, Russia}
\author{H.~Sieber}\affiliation{\it ETH Z\"urich, Institute for Particle Physics and Astrophysics, CH-8093 Z\"urich, Switzerland}
\author{D.~Shchukin}\affiliation{\it P.N.Lebedev Physical Institute of the Russian Academy of Sciences, 119 991 Moscow, Russia}
\author{V.~O.~Tikhomirov}\affiliation{\it P.N.Lebedev Physical Institute of the Russian Academy of Sciences, 119 991 Moscow, Russia}
\author{I.~Tlisova}\affiliation{\it Institute for Nuclear Research, 117312 Moscow, Russia} 
\author{A.~N.~Toropin}\affiliation{\it Institute for Nuclear Research, 117312 Moscow, Russia}
\author{A.~Yu.~Trifonov}\affiliation{\it Tomsk Polytechnic University, 634050 Tomsk, Russia}\affiliation{\it Tomsk State Pedagogical University, 634061 Tomsk, Russia}
\author{B.~I.~Vasilishin}\affiliation{\it Tomsk Polytechnic  University, 634050 Tomsk, Russia}
\author{G.~Vasquez Arenas}\affiliation{\it Universidad T\'{e}cnica Federico Santa Mar\'{i}a, 2390123 Valpara\'{i}so, Chile}
\author{P.~V.~Volkov}\affiliation{\it  Joint Institute for Nuclear Research, 141980 Dubna, Russia}\affiliation{\it Skobeltsyn Institute of Nuclear Physics, Lomonosov Moscow State University, 119991  Moscow, Russia}
\author{V.~Yu.~Volkov}\affiliation{\it Skobeltsyn Institute of Nuclear Physics, Lomonosov Moscow State University, 119991  Moscow, Russia}
\author{P.~Ulloa}\affiliation{\it Departamento de Ciencias F\'{i}sicas, Universidad Andres Bello, Sazi\'{e} 2212, Piso 7, Santiago, Chile}

%
%
\collaboration{The NA64 Collaboration}\noaffiliation
\vskip 0.25cm

\date{\today}

\begin{abstract}
We report the results of a search for a light pseudoscalar particle $a$ that couples to electrons
and decays to $e^+e^-$ performed using the high-energy CERN SPS H4 electron beam. If such light pseudoscalar exists, it 
could explain the ATOMKI anomaly (an excess of $e^+e^-$ pairs in the nuclear transitions of $^8$Be and $^4$He nuclei at the
invariant mass $\simeq 17$ MeV observed by the experiment at the 5 MV Van de Graaff accelerator at ATOMKI, Hungary).
We used the NA64 data collected in the
"visible mode" configuration with a total statistics corresponding to $8.4\times 10^{10}$ electrons on target (EOT) in 2017 and 2018.
In order to increase sensitivity to small coupling parameter $\epsilon$ we used also the data collected in 2016 - 2018 in the
"invisible mode" configuration of NA64 with a total statistics corresponding to $2.84\times 10^{11}$ EOT.
The background and efficiency estimates for these two configurations were retained from our previous analyses searching
for light vector bosons and axion-like particles (ALP) (the latter were assumed to couple predominantly to $\gamma$).
In this work we recalculate the signal yields, which are different due to different cross section and lifetime of a pseudoscalar
particle $a$, and perform a new statistical analysis. As a result, the region
of the two dimensional parameter space $m_a - \epsilon$ in the mass range from 1 to 17.1 MeV is excluded. At the mass of the central 
value of the ATOMKI anomaly (the first result obtained on the berillium nucleus, 16.7 MeV) the values of $\epsilon$
in the range $2.1 \times 10^{-4} < \epsilon < 3.2 \times 10^{-4}$ are excluded.

\end{abstract}

\maketitle

\section{Introduction}
\label{sec:intro}

 Gauge-singlet pseudoscalar particles have been attracting attention for many years in view of understanding the phenomenology of the strong CP
problem (lack of CP violation in QCD) \cite{PecceiQuinn1977,WeinbergAxion1978,WilczekAxion1978}. Such particles appear in models
with a spontaneously broken global symmetry and are considered as candidates for either dark matter or for mediators to a dark sector, see, e.g.,
Refs. \cite{Essig2013dark,alexander2016dark,mb,Beacham:2019nyx,europeanstrategyforparticlephysics2020,LDMXproposal,XENON:2020rca,Buttazzo:2020vfs}.

 Previously, a neutral pseudoscalar particle $a$ decaying to $e^+e^-$ \cite{alves2020signals,Ellwanger:2016wfe} was proposed to explain
the ATOMKI anomaly \cite{be8,be8-2,be8-3}. Such particles could also
cause a deviation from the expected value of the electron anomalous magnetic moment \cite{Parker191,finestructure2020,andreev2021constraints}.

 The NA64 experiment previously derived limits on light vector particles decaying to $e^+e^-$ \cite{visible-2018-analysis}.
The production cross section and decay width of a pseudoscalar particle differ from the corresponding values predicted for a vector
particle with the same mass. In this paper we use the available data of the NA64 experiment and some results of the previous analyses
of these data to derive limits on the particle $a$.

\section{The search method}
\label{sec:method}

\begin{figure*}[tbh!!]
\centering
\includegraphics[width=1.\textwidth]{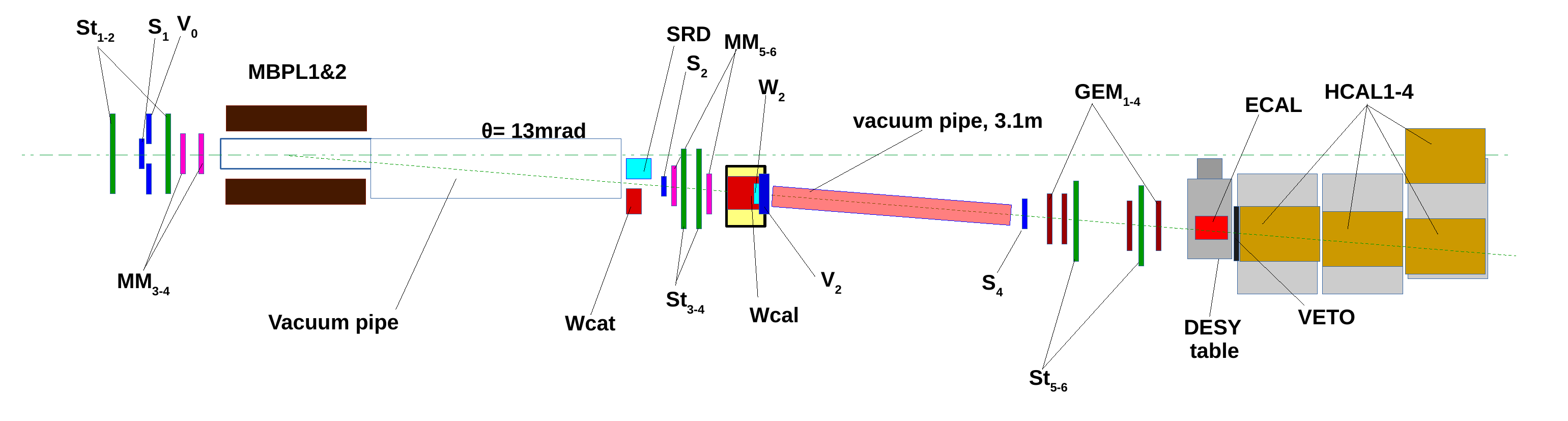}
\caption{The NA64 setup to search for $A'(a)\to \ee$  decays of the bremsstrahlung $A'(a)$ produced in the reaction
$eZ \to eZA'(a) $ of the 150 GeV electrons incident on the active WCAL target. The figure is reproduced from Ref. \cite{visible-2018-analysis}.}
\label{fig:setup}
\end{figure*}

 The NA64 experiment in the "visible mode" configuration, i.e. configured for searches for dark matter particles, such as dark photons A'
or $a$ particles, decaying visibly, into $\ee$ pairs, is described in Refs. \cite{visible-2018-analysis, NA64Be2017} and
shown in Fig.~\ref{fig:setup}.

The experiment uses the high purity H4 electron beam at the CERN SPS (beam energy 100 GeV in 2017 and 150 GeV in 2018).
The backgrounds coming from the beam are further significantly suppressed by using the synchrotron radiation detector (SRD) to identify
electrons \cite{na64-prd}. This suppression factor for the hadron contamination of the beam is $\sim10^{-4}$.
The most important subdetectors in this setup are the two electromagnetic (EM) calorimeters: the compact target-calorimeter WCAL assembled
from the tungsten and plastic scintillator plates with wavelength shifting fiber read-out and ECAL, a matrix of $6\times 6$ shashlik-type
lead - plastic scintillator sandwich modules \cite{na64-prd}. We also use a veto counter $W_2$ placed immediately after the WCAL and
a decay counter $S_4$ installed downstream the vacuum decay tube. Measuring the energy deposition in $W_2$ ensures that no charged particle exits
from the WCAL, while a signal compatible with two Minimum Ionizing Particle (MIP) in $S_4$ indicates that a decay to $e^+e^-$ happened in the decay volume.
The high efficiency thick (5cm) counter VETO and the hadron calorimeter HCAL are installed downstream the ECAL.
The HCAL consists of four modules, three of them are placed at the axis of the beam deflected by the MBPL magnets. They are
usually used as a veto against electroproduction of hadrons in the WCAL. The fourth module serves as a veto against upstream interactions of electrons
before reaching the target. Some most important distances of the setup are shown in Table~\ref{tab:distances}. The distances in the invisible
mode configuration in 2016, 2017 were slightly different, this was taken into account in the exact signal yield calculation, which can be made only
using the detailed simulation.

\begin{table}[tbh!]
\begin{center}
\caption{Some parameters and distances of the NA64 experimental setups.}
\label{tab:distances}
\vspace{0.15cm}
\begin{tabular}{|c|c|c|c|c|}
\hline
Run                        & Beam energy    & Calorimeter size    & Distance                                   & Decay length   \\
                           &                & along the beam      & end of calorimeter{\textendash}end of veto &                \\
\hline
2017 visible mode          & 100 GeV        & 17.3 cm             & 2.7 cm                                     & 3.12 m         \\
\hline
2018 visible mode          & 150 GeV        & 17.3 cm             & 0.6 cm                                     & 3.14 m         \\
\hline
2018 invisible mode geom.  & 100 GeV        & 45 cm               & 198 cm                                     & $\simeq$3.4 m  \\
\hline
\end{tabular}
\end{center}
\end{table}

 If the particle $a$ exists, it would be produced via scattering of high-energy electrons off nuclei of an
active target-dump WCAL due to its coupling to electrons $\epsilon e$, where $e$ is the electron charge
and $\epsilon$ is a coupling parameter \cite{Liu_2017_2}. The Lagrangian term corresponding to the coupling with 
electrons $\psi_e$ is $\mathcal{L}\supset -ie \epsilon a \bar{\psi}_e \gamma_5 \psi_e$.
The $a$ production is followed by its decay into $\ee$ pairs:
\begin{equation}
e^- + Z \to e^- + Z + a   ;~ a \to \ee \,.
\label{ea}
\end{equation}
The $a$ can be detected if it decays in flight beyond the rest of the dump and the veto counter $W_2$ 
in the decay volume. The occurrence of the process (\ref{ea}) would manifest itself as an excess of events with two EM-like
showers in the setup, one in the WCAL and another one in the ECAL, with the total energy $E_\text{tot} = E_{WCAL} +E_{ECAL}$
compatible with the beam energy ($E_0$), above those expected from background sources.
In the design of the "visible mode" setup we took into account that in a
part of the parameter space to be explored the particle $a$ is rather short-lived. The distance from the creation zone
to the end of veto ($W_2$ counter) was minimized.

  \par The candidate events are selected by applying the following main criteria:
\begin{enumerate}
\item
The upper cut on the energy deposition in $W_2$ veto counter is $\sim$0.7 $MIP$ (most probable energy deposition of a minimum ionizing particle),
or 0.0007 GeV;
\item
The lower cut on the signal in $S_4$ decay counter is 1.5 $MIP$s (0.0003 GeV);
\item
$E_0 - E_\text{tot}$ is smaller than double total uncertainty of this difference; the energy in the downstream calorimeter $E_{ECAL} > $ 25 GeV;
\item
The shower in the WCAL must be consistent with that produced by a primary electron, we use a WCAL pre-shower lower-energy cut of 0.5 GeV to check this;
\item
The cell with maximal energy deposition in the ECAL should be the one on the deflected beam axis;
\item
The longitudinal and lateral profiles of the shower in the ECAL are consistent with a single EM shower. The longitudinal shape
is checked by requiring an energy deposition of at least 3 GeV in the ECAL pre-shower. The lateral profile of the shower was compared to
the profile measured in the calibration beam using the $\chi^2$ method. This does not decrease the efficiency
for signal events because the distance between $e^-$ and $e^+$ in the ECAL is significantly smaller than the ECAL cell size of 3.82$\times$3.83 cm$^2$ ;
\item
The rejection of events with hadrons in the final state is based on the energy deposition in the VETO counter (less than 0.9 $MIP$ = 0.09 GeV required)
and the hadron calorimeter HCAL (less than 1 GeV for each module required).
\end{enumerate}

 The cuts used for the event selection are explained in more details in the previous paper \cite{visible-2018-analysis}.

 In order to increase the sensitivity to $a$ at small values of $\epsilon$ (below $\sim 2 \times 10^{-4}$), we also used
the NA64 data collected in 2016 - 2018 in the "invisible mode" configuration \cite{Banerjee_2019} with only one electromagnetic calorimeter ECAL
serving as a target, with an analysis scheme exactly as in our ALP search \cite{Banerjee-ALP-2020} (the picture of the setup can be found in
the same reference).
In this method the HCAL is used not only as a veto, but also as a detector of possible $a \rightarrow \ee$ decays.


 The "invisible mode" configurations are characterized by much longer distance from the creation zone to the end of veto, as can be seen
in Table~\ref{tab:distances}. However, for small values of $\epsilon$ this is not a problem as the particle $a$ is relatively long-lived.
There is a significant probability that after
its creation in the ECAL and passing the first HCAL1 module serving as a shield/veto it would be observed in the NA64 detector in one of
the two signatures: (S1) as an event with $a \rightarrow \ee$ decay inside the HCAL2 or HCAL3 modules (HCAL2,3 in the following), or (S2) as an event
with a significant missing energy if it decays beyond HCAL2,3.
In both cases the main requirements were that the
shower profile in ECAL is compatible with electron, the VETO counter signal is smaller than 0.9 $MIP$ and that the energy deposition
in HCAL1 is smaller than 1 GeV. The main requirements for the signature (S1) event were that the total energy deposition in HCAL $E_{HCAL}\gtrsim 15$ GeV,
and that the energy deposited in HCAL2,3 is concentrated in the central cell \cite{Banerjee-ALP-2020}. For the signature (S2) the total
energy deposition in ECAL was required to be smaller than 50 GeV and the energy in all HCAL modules should be smaller than 1 GeV.
There was also a number of other criteria explained in more details for the signature (S1) in \cite{Banerjee-ALP-2020} and
for the signature (S2) in \cite{Banerjee_2019,na64-prd}.

 As the event selection was exactly the same as in the previous analyses, we reused the results of the background estimation from them.
The main background in the NA64 "visible mode" configuration comes from the electroproduction of $K^0_S$ and their decays
$K^0_S \rightarrow \pi^0 \pi^0$ in flight, followed by conversion of one of the decay photons. After optimization of the setup in 2018 this
background, determined from data, amounted to less than 0.005 events per $10^{10}$ EOT \cite{visible-2018-analysis}. The main background
in the "invisible mode" configuration comes from neutral hadron production by electrons in the target. These neutral hadrons either pass without
interaction the first HCAL module and deposit energy in the downstream modules HCAL2,3, or completely escape detection because of insufficient
aperture of the HCAL. These backgrounds, of the order of 0.1 events, were also determined from data \cite{Banerjee-ALP-2020,Banerjee_2019}.

\section{Signal yield and results}
\label{sec:results}

 In the calculations of the signal yield we used the fully Geant4 \cite{geant4} compatible package DMG4 \cite{DMG4_paper1}.
This package can simulate the production of four types of DM mediator particles in the electron bremsstrahlung processes, including
the vector and pseudoscalar cases. It contains a collection of corresponding cross sections, total and differential, including the 
ones for a pseudoscalar particle $a$ from the model of Ref. \cite{alves2020signals}. The total cross sections are calculated at
the exact tree level (ETL). We assumed that the $a$ decay branching ratio to $e^+e^-$ is 100\%.

 The package DMG4 was compiled together with the program based on Geant4 for the full simulation of the NA64 experimental setup.
The produced signal samples were processed by the same reconstruction program as the real data and passed the same selection
criteria.

 We remind that no candidate events were found in all previously made analyses that we reuse and combine here.
For the statistical analysis, there were three main data bins, see Table~\ref{tab:bins}.
The bins 1 and 3 were further subdivided into bins corresponding to different years and conditions. The total number of bins was up to 9.
The backgrounds and various uncertainties in these bins were estimated in the previously published
analyses \cite{visible-2018-analysis,Banerjee-ALP-2020,Banerjee_2019,na64-prd} and reused.
This concerns also most of the signal yield uncertainties. The uncertainties depending on the $a$ energy and path to decay
distributions were recalculated for the new signal samples, but turned out to be compatible with the values determined previously
and remained unchanged. All uncertainties, summed up in quadrature, don't exceed 20\%.

\begin{table}[tbh!]
\begin{center}
\caption{Main data bins of the statistical analysis.}
\label{tab:bins}
\vspace{0.15cm}
\begin{tabular}{|c|c|c|c|}
\hline
Data bin                                & 1. Vis. mode config. & 2. Invis. mode config., signature S1 & 3. Invis. mode config., signature S2 \\
\hline
Total number of EOT                     & $8.4\times 10^{10}$  & $2.84\times 10^{11}$                 & $2.84\times 10^{11}$                 \\
\hline 
Background                              & 0.083                & 0.1                                  & 0.472                                \\
\hline
Signal $m_a$=10 MeV, $\epsilon=10^{-4}$ & 1.32                 & 6.7                                  & 1.4                                  \\
\hline
\end{tabular}
\end{center}
\end{table}

\begin{figure*}[t]
\begin{center}
\includegraphics[width=0.8\textwidth]{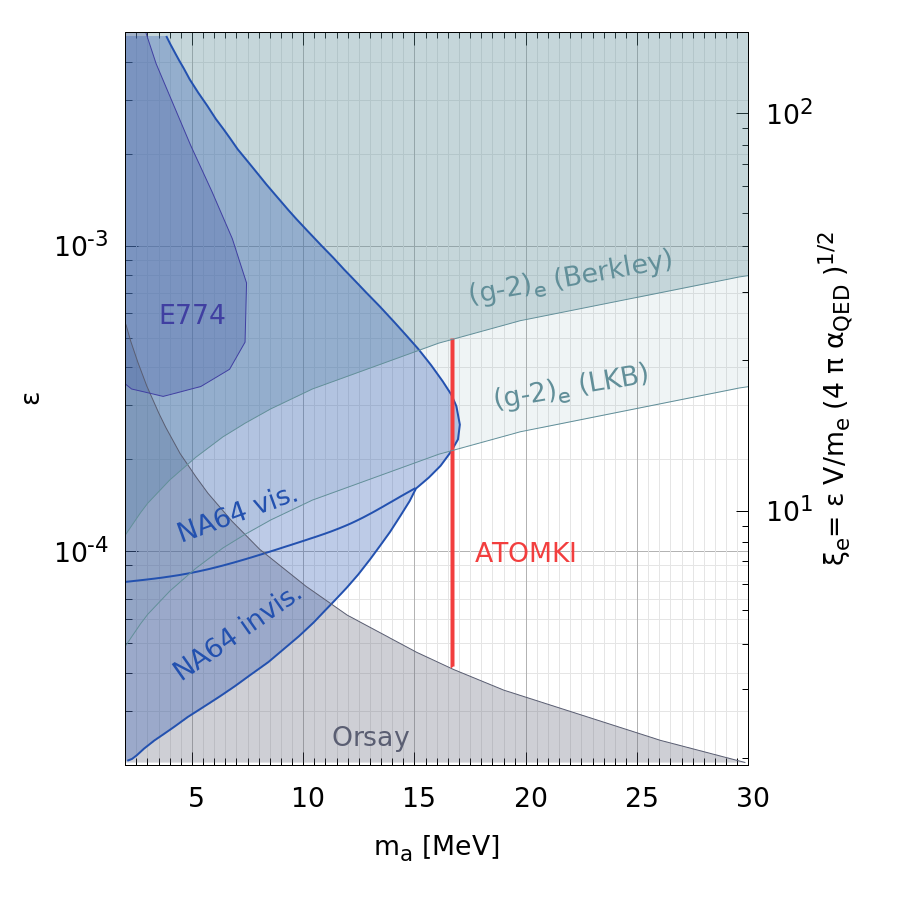}
\caption {The 90\% C.L. limits on the pseudoscalar particles decaying to $e^+e^-$ pairs.
          On the right vertical axis we use the standard notation for the pseudo-scalar coupling $\xi_e=\epsilon (V/m_e) \sqrt{4 \pi \alpha_{QED}}$,
          where $V=246$~GeV is a vacuum expectation value of the Higgs field \cite{Andreas:2010ms}. This corresponds to the Lagrangian term
          $\mathcal{L}\supset -i \xi_e \frac{m_e}{V} a \bar{\psi}_e \gamma_5 \psi_e$.
          The red vertical line corresponds to the ATOMKI anomaly at $m_a = 16.7$ MeV (central value of the first result on berillium).
          The $\epsilon$ range excluded at this mass is $2.1\times 10^{-4} < \epsilon < 3.2\times 10^{-4}$.
          The region excluded using only the data collected with the visible mode geometry is denoted as "NA64 vis.", the extention
          of this region obtained using all data is denoted as "NA64 invis.".
          The regions excluded by the $(g-2)_e$ measurements (Berkley \cite{Parker191} and LKB \cite{finestructure2020}) are shown.
          The limits from the electron beam-dump experiments E774~\cite{bross} and Orsay~\cite{dav} are taken from Ref.~\cite{Andreas:2010ms}.
\label{fig:result}
}
\end{center}
\end{figure*}

 The exclusion limits were calculated by employing the multi-bin limit setting technique in a program based on
RooStats package \cite{root-roostats} with the modified frequentist approach, using the profile likelihood
as a test statistic \cite{Junk_1999,STATIS,Read:2002hq}.
The 90\% C.L. excluded region in the two-dimensional plot $m_a - \epsilon$ is shown in Fig.~\ref{fig:result}.
The regions excluded by the $(g-2)_e$ measurements are also shown, the most stringent is LKB \cite{finestructure2020}.
The central value of this measurement has the sign opposite to possible contribution from a pseudoscalar particle $a$ coupled to electrons.
We used a frequentist approach to calculate the 90\% C.L. limit from it. 
We note that the limits from the $(g-2)_e$ measurements are model-dependent and can be significantly less strict in some scenarios
\cite{Buttazzo:2020vfs,Alves:2017avw}.

\section{Conclusion}

 We performed a model-independent search for light pseudoscalar particles that couple to electrons and decay predominantly
to $e^+e^-$ pairs in the NA64 experiment at the CERN SPS North Area. The active target-calorimeter of this experiment was exposed
to the electron beams with the energy of 100 and 150 GeV. No signal of such particles was found, allowing us to exclude the region
of the $(m_a, \epsilon)$ parameter space in the mass range from 1 to 17.1 MeV. Additional exposure will increase sensitivity,
in particular at the mass of the ATOMKI anomaly of 16.7 MeV \cite{depero2020hunting}.

\section{Aknowledgements}

 We gratefully acknowledge the support of the CERN management and staff and the technical staff of the participating institutions
for their vital contributions. We thank Robert Ziegler for useful comments about the $a$ contribution to $(g-2)_e$.
 This work was supported by the Helmholtz-Institut f\"ur Strahlen- und Kern-physik (HISKP), University of Bonn, 
the Carl Zeiss Foundation 0653-2.8/581/2, and  Verbundprojekt-05A17VTA-CRESST-XENON (Germany), Joint Institute for Nuclear Research (JINR) (Dubna), 
the Ministry of Science and Higher Education (MSHE) and RAS (Russia), ETH Zurich and SNSF Grant No. 169133, 186181 and 186158 (Switzerland), 
Tomsk Polytechnic University, FONDECYT Grants No.1191103, No. 190845 and No. 3170852, UTFSM PI~M~18~13, ANID PIA/APOYO AFB180002 (Chile).
The work on the DMG4 package was supported by the Russian Science Foundation RSF grant 21-12-00379.

\bibliography{bibliographyOther,bibliographyNA64,bibliographyNA64exp}
\bibliographystyle{na64bib}
\bibliographystyle{na64-epjc}

\end{document}